# One-dimensional mapping of femtosecond laser filaments using coherent microwave scattering

Nicholas Babusis[1], Adam Patel[1], Rokas Jutas[2], Zahra Manzoor,[1] Mikhail N. Shneider,[3] Audrius Pugzlys[2], Andrius Baltuska[2], and Alexey Shashurin[1]

[1] *School of Aeronautics and Astronautics, Purdue University, West Lafayette, IN, USA*
[2] *Photonics Institute, Vienna University of Technology, Vienna, Austria*
[3] *Mechanical and Aerospace Engineering, Princeton University, Princeton, NJ, USA*

**Abstract**

This paper reports on the use of coherent microwave scattering (CMS) for spatially resolved electron number density measurements of elongated plasma structures induced at mid-IR femtosecond filamentation in air. The presented studies comprise one-dimensional mapping of laser filaments induced via 3.9 μm, 127.3 fs laser pulses at output energies up to 15 mJ. The axial electron number density was measured to be invariant (about $2\times10^{15}$ cm$^{-3}$) along the entire filament length and for all tested laser pulse energies 5-15 mJ, and the corresponding laser intensity in the middle portion of the filament was estimated to be nearly constant for 5-15 mJ pulse energies (about 30-40 TW/cm$^2$). These fundings support that intensity clamping conditions were achieved in the experiments. The proposed approach enables capabilities that are currently unavailable to perform absolute and longitudinally resolved measurements of electron number density in laser filaments and to precisely characterize conditions associated with self-focusing and intensity clamping.

## Introduction

Field ionization in gases by intense femtosecond laser pulses have enabled many frontier fundamental research topics in nonlinear optics such as high-order harmonic generation, filamentation, and standoff nonlinear spectroscopy of the atmosphere.[1,2,3,4] However, measurements of electron number densities in laser filaments are particularly challenging. Multiple diagnostic methods have been evaluated with different degree of success, including optical interferometry,[5,6] shadowgraphy,[7] time-of-flight mass spectrometry,[8,9] capacitive probes,[10,11,12] scattering of THz radiation,[5,6] avalanche electron multiplication,[13] attenuation of rectangular waveguide mode.[14] Overall, these methods are characterized by limited range of applicability, ambiguous interpretation, being indirect and semiempirical (require absolute calibration based upon theoretically predictions), so that plasma diagnostics of laser-induced photoionization remains a challenging task.

Measurements of the electron number densities in the mid-IR laser filaments in air are especially challenging as they are associated with relatively low plasma densities in comparison to the near-IR counterparts due to the stronger action of the plasma defocusing term in the nonlinear refractive index at larger wavelength ($\frac{\omega_p^2}{2\omega^2}$, where $\omega_p$ is the plasma frequency and $\omega$ is the laser frequency). Numerical simulations predict $n_e$ in the range $10^{15}$-$10^{16}$cm$^{-3}$ for ~100 fs laser pulses at 20-30 mJ focused using 45-200 cm focal length lens.[15,16] At the same time, experimental data on the electron number density in the mid-IR laser filaments are scanty and controversial. Inline holographic microscopy method yields questionable $n_e$ nearly 3-4 orders of magnitude larger than that predicted theoretically.[17] Data obtained from capacitive probes and digital photography methods provide only relative measurements and lack absolute calibration.[10,17]

Recently, we utilized the method of coherent microwave scattering (CMS) to measure electron number densities and photoionization rates of laser-induced photoionization in mid-IR.[18] The electron number densities in the range $10^{15}$–$1.8\times10^{16}$ cm$^{-3}$ (and corresponding photoionization rates $5.0\times10^8$–$6.1\times10^9$ s$^{-1}$) were reported for intensities in the range 13-190 TW/cm$^2$, respectively. However, these CMS measurements were conducted for relatively low fs-laser pulse energies (≤3.5 mJ) when the effect of Kerr and plasma nonlinear terms of the refractive index on beam propagation is negligible, and nearly Gaussian beam can be safely assumed in vicinity of the beam waist. At the same time, conditions associated with strong optical nonlinearities and substantial perturbations of the beam profile from the Gaussian beam (e.g., intensity clamping, self-focusing)[3,19,20] possess significant practical interest and require adequate experimental evaluation. In this work we further develop CMS technique to conduct spatially resolved measurements of electron number densities along mid-IR laser filaments in air. To incorporate some level of spatial resolution, these filaments were enclosed in two conductive tubes with an exposed central air gap. The proposed technique enables unprecedented, currently unavailable capabilities to conduct absolute and spatially resolved measurements of electron number density in laser filaments and to precisely characterize conditions associated with self-focusing and intensity clamping.

# Methodology

In this experiment, a mid-IR femtosecond laser system ($\lambda_L = 3.9$ μm, $\tau_{FWHM} = 127.3$ fs, pulse energy up to 30 mJ) was utilized.[15] The laser was focused in air at variable axial positions by a $CaF_2$ lens (Thorlabs LA5956-E, $f$=750 mm at 588 nm wavelength) on a linear stage. The full width at half maximum (FWHM) pulse duration was measured via second-harmonic generation frequency-resolved optical gating (SHG FROG). The $1/e^2$ beam radiuses prior the lens were measured using a beam profiler to be 4.02 and 4.20 mm in the horizontal and vertical directions respectively. Laser pulse energies between 5 and 15 mJ were used in this work as measured via a pyroelectric energy meter before the lens.

In the CMS circuit, shown schematically in Figure 1(a), 11 GHz microwaves generated by the Tx horn irradiated the filament (linearly polarized along the filament axis). The +38 dBm power radiated by the Tx horn created an incident power density of about 10 mW/cm² at the beam axis, which is sufficiently low to prevent thermal disturbance of the plasma object by probing microwaves.[21,22] Most of the filament was contained inside a shield tube enclosure with a 15-mm exposed central air gap, as discussed in details in the Results and Discussion section. The Rx horn, azimuthally offset by 90º from the Tx horn, as shown in Figure 1(b), received microwave radiation scattered from the exposed plasma segment. The Rx and Tx horns were placed 190 mm and 240 mm from the beam axis respectively and have microwave-absorbent backdrops. The scattered signal was then detected using an I/Q mixer (HMC-C042), and the low frequency I/Q signal outputs were amplified by the post-mixer amplifiers equipped with DC blocks (PE15A1007 amplifier with 9 kHz - 3 GHz bandwidth) to eliminate contribution of the environmental reflections to the output signal. Calibrated amplification of the MW scattering signals and of the I/Q channels were used as needed to improve system sensitivity, and linear, undistorted amplifier operation was ensured throughout experiments. A phase shifter was added to the transmitting arm for easy monitoring and confirming that the plasma scattered signal phase is consistent with the collisional scattering regime (90 degrees phase shift between the electron displacement and the incident electric field)[21] throughout experiments. The response time of the CMS system was ~0.5 ns, governed by the 2 GHz bandwidth of I/Q mixer output. A Rohde & Schwarz RTO 2044 oscilloscope with 4 GHz bandwidth was used for the signal recording.

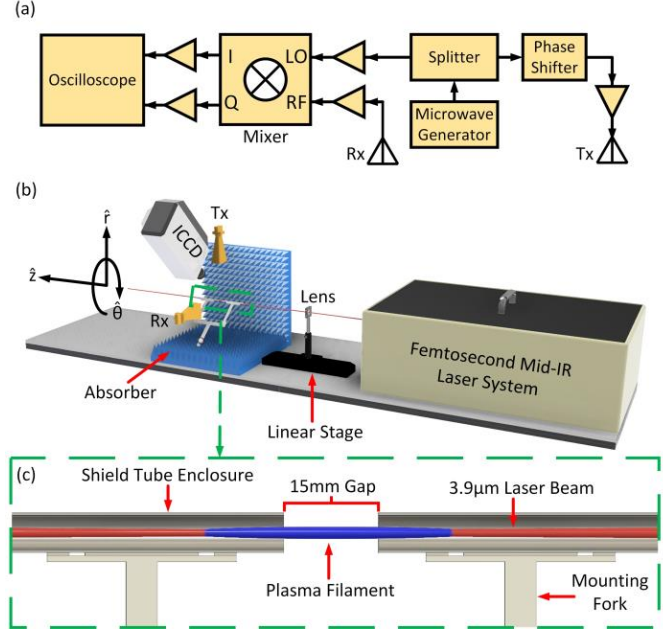

**Figure 1:** Schematics of the experimental setup containing the 3.9-μm femtosecond laser and coherent microwave scattering system equipped with metal tube enclosure for spatially resolved measurements.

## Results and Discussion

A long-exposure representative photograph of the considered laser filaments is shown in Figure 2. The parameters of the laser system and focusing lens considered in this work restrict the filament length to a few centimeters. Note that, for the laser pulse energies used here ($E_L$=5-15 mJ), nonlinear effects of refractive index $n = n_0 + n_2 I_L - \frac{\omega_p^2}{2\omega_L^2}$ (Kerr and plasma nonlinear terms - $n_2 I_L$ and $\frac{\omega_p^2}{2\omega_L^2}$, respectively) were significant, where $n$ is the nonlinear index of refraction, $n_0$ is the linear component of the index of refraction, $n_2$ is the coefficient of the Kerr nonlinear term, $I_L$ is the laser intensity, $\omega_L$ is the laser field frequency, $\omega_p = \sqrt{\frac{e^2 n_e}{\varepsilon_0 m}}$ is the plasma frequency, $e$ is the electron charge, $\varepsilon_0$ is the dielectric permittivity of vacuum, and $m$ is the electron mass. The critical power of self-focusing for 3.9 μm wavelength can be roughly estimated as $P_{cr} = \frac{3.72 \lambda^2}{8\pi n_0 n_2} = 45\ GW$.[23] The corresponding laser pulse energy can be estimated as $P_{cr} \frac{\tau_{FWHM}\sqrt{\pi}}{2\sqrt{\ln 2}} = 6.1$ mJ for the considered here laser system, which is well below the maximum examined laser pulse energy of 15 mJ.

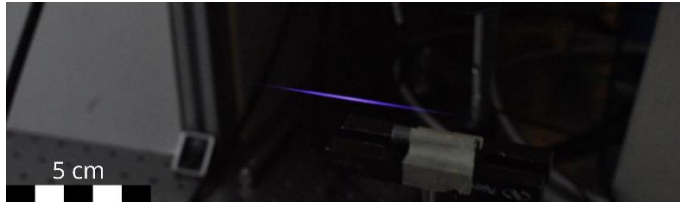

**Figure 2:** Representative photograph of mid-IR laser filaments initiated in atmospheric air with a 750 mm focal length lens.

The standard CMS technique was modified in this work to produce 1-D spatially resolved $n_e$ measurements by shielding most of the plasma filament inside the metal shield tube enclosure shown in Figure 1(c). Only plasma in the gap between the tubes is exposed to the microwave radiation, and, therefore, only plasma in the gap is being diagnosed. The shield tubes (OD: 6.35 mm, ID: 4.57 mm) were separated by a microwave-accessible gap of 15 mm, or approximately half wavelength of the probing microwave radiation. The tubes were supported by a mounting fork made of PLA plastic to minimize reflections. The exposed plasma segment was irradiated by the microwaves from the Tx horn, which were scattered off the plasma and detected by the Rx horn. The tube enclosure was positioned with CMS horns in the center-gap plane while the filament position was adjusted by translating the lens along the z-axis to acquire measurements of different portions of the filament. The length of the shield tubes was selected to ensure that the ~10-40 mm long plasma filaments can be fully contained within either tube.

The first step in testing this new method was to confirm minimal disruption of the filament characteristics due to the tube enclosure. We specifically restrict this non-intrusiveness requirement to the period <10 ns after the laser pulse when CMS data was acquired. On this fast time scale, any hydrodynamic perturbations due to the presence of the tube can be neglected. The minimum distance between the plasma filament and tube enclosure was 0.86 mm (accounting for a ≤ 1 mm alignment error). The corresponding minimum time for hydrodynamic interactions, estimated as the time for a shock wave (generated by the filament) to travel back and forth across this distance, is ~5 μs, which is much longer than CMS observation time of ~10 ns.

We experimentally verified that the tubes are nonintrusive for the filament properties right after the laser pulse by analyzing and comparing the visible light emissions of shielded and unshielded filaments. Figure 3 depicts intensified charge coupled device (ICCD) imaging of plasma emissions from shielded and unshielded filaments taken with an exposure time of 100 ns and averaged over 15 laser shots. The bandwidth of the imaging system comprised of Princeton

Instruments PI-Max 4 ICCD camera and Micro-Nikkor 55mm lens was 375-900 nm (the observed filament images are likely dominated by spectral lines of excited and ionized nitrogen molecules, which are expected to closely correlate with the electron number density).[17,24,25] For each laser pulse energy $E_L$, the upper plot shows images with the tube enclosure present (stitched together at the edge of the exposed region, indicated by the red lines), and the lower plot shows an image taken without the tube enclosure present. Visually, the invariance of plasma emission features implies minimal perturbation of the filament from the tube enclosure. One can further observe enhanced Kerr-focusing at increased $E_L$ indicated by the shifting of the filament centroid toward the laser (z-axis in Figure 3 indicates the distance from the focusing lens). Note that the filament was observed at slightly larger distances than 750 mm due to wavelength-dependent focal length shift of Thorlabs LA5956-E lens and slight divergence of the beam prior to the lens.

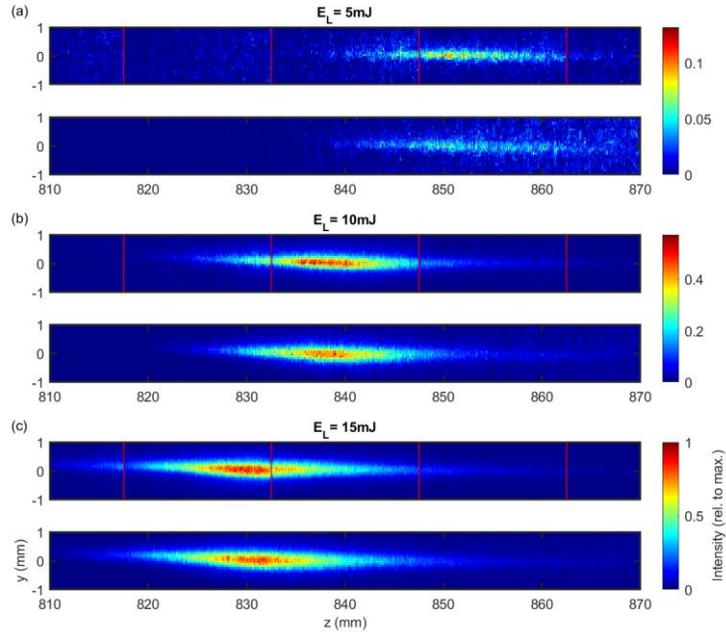

**Figure 3:** ICCD Images of mir-IR laser filaments with (upper) and without (lower) tube enclosure present. Red lines indicate boundaries of stitched images when tube enclosure is present.

The next step was to modify a standard procedure of evaluating total electron count in the tested plasma object $N_e$ from the measured scattered signal $V_S$ (developed and widely utilized previously for conventional CMS technique),[21,26,27,28] so that it accounts for presence of the tube enclosure. Presence of the tube enclosure leads to a reduction of electric field magnitude in the gap due to partial shielding by the tubes, so that tested plasma is exposed to a weaker electric field, and perturbs phase of the field in vicinity of the tube ends, so that constructive interference between wavelets at the Rx horn is partially disturbed.[21] Both effects lead to reduction of the scattered signal measured at the Rx horn. Therefore, the presence of the tube enclosure was accounted via incorporation of the scattered signal reduction factor ($\kappa$). To quantify this reduction factor experimentally, the scattered signal $V_S$ was measured for about 15 mm long plasma filament ($E_L$ = 2.8 mJ) placed in the center of the 15 mm gap of the tube enclosure and without the tube enclosure. It was found that $V_S$ reduced by the factor $\kappa = 4.85 \pm 0.9$ in the presence of the tube enclosure as illustrated in Figure 4. Finally, the scattering signal reduction factor $\kappa$ was incorporated into the final equations relating $V_s$ and $N_e$ as follows:

$$V_s = \begin{cases} AV_D\epsilon_0(\epsilon_D - 1)\omega & - \text{ dielectric object} \\ \dfrac{A}{\kappa}\dfrac{e^2}{m\nu_m}N_e & - \text{ plasma object in tube enclosure} \end{cases} \quad \textbf{Eq. 1}$$

where $A$ is the calibration factor determined via standard dielectric scatterer calibration procedure (free-space, no tube enclosure) described elsewhere, [21,26,27,28] $\epsilon_0$ is the vacuum permittivity, $\epsilon_D$ is the dielectric relative permittivity, and $V_D$ is the dielectric object volume. Note that the collisional scattering case ($\nu_m\omega \gg |\xi\omega_p^2 - \omega^2|$) was satisfied for all experiments reported in this work.[21]

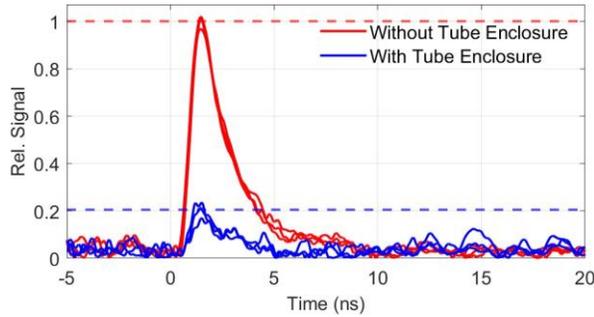

**Figure 4:** Reduction of microwave scattering signal in the presence of the tube enclosure.

Using the procedure described above and Eq. 1, the total electron number in the exposed portion of the filament ($N_e$) was determined, as shown in Figure 5. $N_e$ was determined from the measured scattered signal $V_s$ using Eq. 1 and assuming a constant collisional frequency of $\nu_m$=2×10$^{12}$ s$^{-1}$ corresponding to the electron temperature of $T_e$ ~ 1 eV expected on the timescale of ~1 ns after the laser pulse.[5,18,21,25,29,30] As expected, the measured signal is temporally characterized by a rapid increase of $N_e$ due to laser-induced photoionization of air followed by an electron decay (twofold decay over ~2 ns). The electron decay is governed by dissociative recombination ($X_2^+ + e \rightarrow X + X$, where $X$ is $N$ or $O$) and three-body attachment to oxygen ($e + O_2 + X_2 \rightarrow O_2^- + X_2$, where $X$ is $N$ or $O$).[5,18,21,25] Comparable contributions of the attachment and dissociative recombination are expected on the initial stage of decay in the central portion of the filament where electron number density was about $2\times10^{15}$ cm$^{-3}$ (reported later in this manuscript). Specifically, using known dissociative recombination coefficient for nitrogen and oxygen ($\beta = 2 \times 10^{-7} \times \sqrt{\frac{300}{T_e}}$ [cm$^3$s$^{-1}$]) and three-body attachment constants for air constituents ($k_{O_2} = 1.4 \times 10^{-29} \times \frac{300}{T_e} \times \exp\left(-\frac{600}{T_{gas}}\right) \times \exp\left(\frac{700\times(T_e-T_{gas})}{T_e \times T_{gas}}\right)$ [cm$^6$s$^{-1}$] and $k_{N_2} = 1.07 \times 10^{-31} \times \left(\frac{300}{T_e}\right)^2 \times \exp\left(-\frac{70}{T_{gas}}\right) \times \exp\left(\frac{1500\times(T_e-T_{gas})}{T_e \times T_{gas}}\right)$ [cm$^6$s$^{-1}$]),[31] one can evaluate comparable rates of dissociative recombination $\beta n_e \sim 1.5\times10^8$ s$^{-1}$ and attachment to oxygen $\nu = k_1 n_{O_2}^2 + k_2 n_{O_2} n_{N_2} \sim 0.7\times10^8$ s$^{-1}$ for the initial stage of decay. On the later stage, as the electron number density decreases, the electron decay is expected to be dominated by the attachment to oxygen.[32] Note that we present only a brief analysis of the electron decay as it is beyond the focus of the current study.

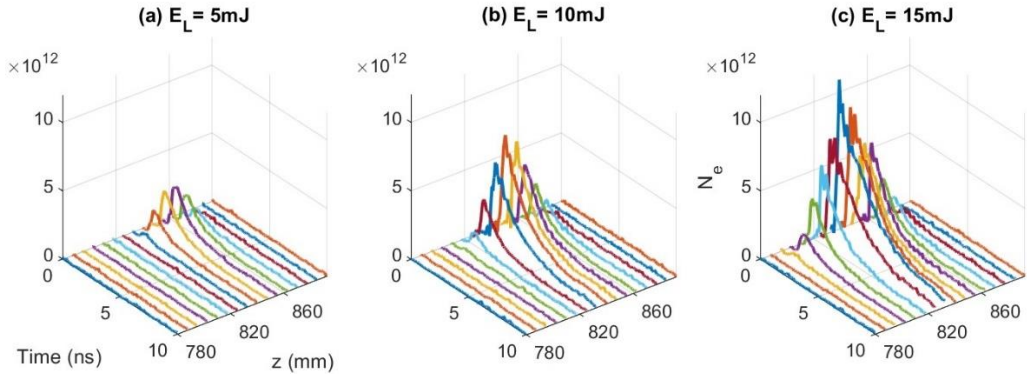

**Figure 5:** Total electron count in 15-mm-long exposed sections of laser filaments.

Total numbers of electrons in the exposed 15-mm-long sections of the filament immediately after the laser pulse ($N_{e0}$) were derived from the peak values of temporal evolutions of $N_e$ shown in Figure 5. The corresponding distributions of $N_{e0}$ along the filament length are presented in Figure 7(a). With the 0.5 ns response time of the CMS system used here and the slower (~2ns) observed $N_e$-decay time, it is expected that $N_{e0}$ reasonably well estimates the total number of electrons produced by the laser pulse.

To convert the total electron number $N_{e0}$ produced by the laser pulse into a corresponding measurement of electron number density, the shape of the radial $n_e$ profile was first evaluated from ICCD imaging as follows. The radial $n_e$ profiles were approximated from the corresponding radial plasma emission profiles, which were evaluated from the ICCD images of the plasma channels shown in Figure 3. Typical radial profiles of emission of the plasma channel, obtained directly from the line-of-sight ICCD images illustrated in Figure 3, are shown in Figure 6. One can see that these profiles can be reasonably well approximated by the Gaussian shape. Therefore, the characteristic plasma radius can be determined directly from the line-of-sight plasma emission profiles and no Abel inversion is required (as Abel inversion of a Gaussian profile retains the same shape and characteristic radius). Thus, plasma emission profiles shown in Figure 6 were approximated by the Gaussian profile $\left(\propto e^{-\frac{2\,r^2}{r_p^2}}\right)$, and, finally, the same radial profiles of the electron number density were assumed: $n_e \propto e^{-\frac{2\,r^2}{r_p^2}}$. Note that, in agreement with previously reported results, FWHM diameter of the plasma channel of 3.9-μm filaments (e.g., ~500 μm at 15 mJ according to Figure 6) was substantially larger than that of 800-nm filaments (~100 μm), due to stronger action of plasma defocusing in mid-IR.[15,16,33]

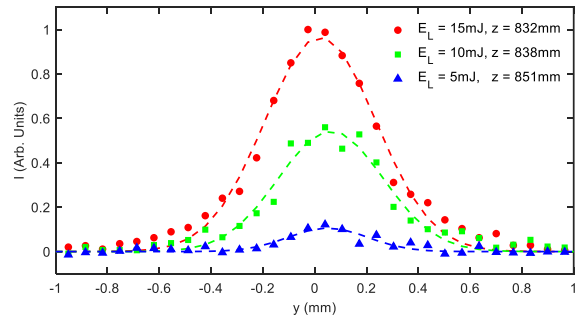

**Figure 6:** Characteristic line-of-sight plasma emission profiles measured by ICCD camera.

The characteristic plasma radius $r_p$ evaluated using the procedure described above is shown in Figure 7(b). The figure illustrates variation of $r_p$ (averaged over each 15-mm-long sections) along the length of the plasma channel (z-axis). Regions associated with low plasma emission (signal/noise ratios ≤ 1.7) at the filament ends were disregarded.

Finally, the electron number density on the axis ($r = 0$), immediately after the laser pulse $n_{e0}(z)$ was determined from the longitudinal distribution of $N_{e0}(z)$ shown in Figure 7(a) and the characteristic plasma radius $r_p(z)$ shown in Figure 7(b) using the following equation:

$$N_{e0}(z) = \iint n_e(r',z') 2\pi r' dr' dz' = \int_{z-\frac{h}{2}}^{z+\frac{h}{2}} \int_0^\infty n_{e0}(z') e^{-\frac{2 r'^2}{r_p^2}} 2\pi r' dr' dz'$$

$$= \int_{z-\frac{h}{2}}^{z+\frac{h}{2}} n_{e0}(z') \frac{\pi r_p^2(z')}{2} dz' \approx n_{e0}(z) \frac{\pi r_p^2(z)}{2} h$$

Eq. 2

Note that variations of plasma parameters ($r_p$ and $n_{e0}$) along the 15 mm gap were assumed to be small, so that the corresponding properties were approximated by their values in the gap center. On the practical level, the gap size should be selected small enough so that variations of plasma parameters along it can be neglected.

The distribution of electron number density along the filament axis $n_{e0}(z)$ is shown in Figure 7(c). One can see that relatively constant value of $n_{e0}$~$2\times10^{15}$ cm$^{-3}$ along the axis was determined for the tested laser pulse energies 5-15 mJ. These measurements demonstrate feasibility of the CMS technique utilized here for absolute and longitudinally resolved measurements of electron number density in laser filaments. Note that the current approach, in contrast with our recent work,[18] does not assume negligible nonlinear effects of refractive index and can be utilized to monitor parameters of laser filaments well into the nonlinear regime, when simultaneous action of Kerr and plasma nonlinearities is significant.

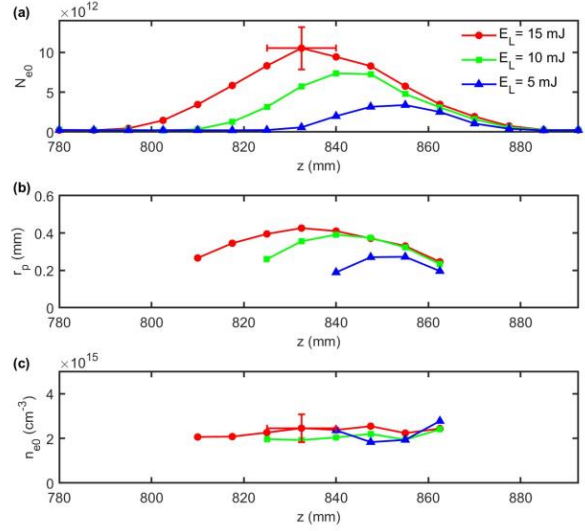

**Figure 7:** (a) Total electron numbers in the 15-mm long segments of the laser filament for various $E_L$; (b) Characteristic plasma radius evaluated from radial plasma emission profiles using ICCD imaging; (c) Corresponding distribution of axial electron number density along the filament.

The observed results can be interpreted as the early stage (onset) of the intensity clamping (associated with balance of nonlinear optical Kerr terms $(n_2 I_L + n_4 I_L^2 + \cdots)$ and nonlinear plasma term $\left(\frac{\omega_p^2}{2\omega^2}\right)$ in the refractive index of air). Indeed, Figure 7(c) indicates that axial number density $2\times 10^{15}$ cm$^{-3}$ was nearly invariant with respect to axial position over the central filament region or laser pulse energy change between 5 and 15 mJ. In addition, the laser intensity in middle section of the filament was also estimated to be nearly constant in the range of about 30-40 TW/cm$^2$ for the tested laser energies 5-15 mJ, which further supports the intensity clamping conclusion (and is consistent with other recent works).[16,17,34] Note that the above intensity values (30-40 TW/cm$^2$) were estimated assuming nearly Gaussian beam intensity distribution in spatial and temporal domains in the vicinity of the middle section of the filament using $I = \frac{E_L}{\frac{1}{2}\pi r_0^2} \frac{2\sqrt{\ln 2}}{\tau_{FWHM}\sqrt{\pi}}$, where the beam radius $w_0$ equal to the plasma radius $r_p$ was used for the middle of the plasma channel. This assumption is justified by nearly Gaussian radial profile of $n_e$ as described above and by reasonable speculation that nearly the entire laser energy is concentrated within $r_p$ in the middle section of the filament (e.g., at z=830 mm for 15 mJ). The latter is based on the observation of two beam waists associated with early focusing of the inner part of the beam and subsequent focusing of the outer part of the beam (e.g., at z=810 mm and 862 mm for 15 mJ). Thus, the substantial portion of the laser pulse energy might be concentrated on the radial periphery beyond $r_p$ in vicinity of the beam waists (e.g., at z=810 mm and 862 mm for 15 mJ); however, concentration of nearly entire laser beam energy within $r_p$ is expected in the middle portion of the filament (e.g., at z=830 mm for 15 mJ).

Several comments and refinements can be recommended for future work. First, although conventional filaments often persist for meters due to the competition between intensity-clamping and plasma / diffraction-based defocusing, the filaments considered here were fairly short (up to ~3-4 cm) as the utilized laser system operated near the very onset of the intensity clamping. In future work, these measurements should be conducted with more conventional, longer laser filaments (~10-100 cm) similar to those reported recently.[15,16,17] Second, the method proposed here is associated with a substantial level of reflected power from the metal tube enclosure contributing to larger background noise and correspondingly decreasing the signal to noise ratio. To eliminate the need for shielding any part of the plasma object by the metal tube enclosure, the use of a focused smaller-wavelength microwave beam may be investigated for $n_e$-measurements spatially resolved by the microwave beam's diameter (e.g., ~ 30 GHz focused microwave beam would enable spatial resolution of ~1 cm).

## Conclusion

This paper demonstrates feasibility of the use of coherent microwave scattering technique for absolute and longitudinally resolved measurements of electron number density in laser filaments in conditions associated with self-focusing and intensity clamping. The technique can be further improved by using focused smaller-wavelength microwave beams instead of metal tube enclosure.

## Acknowledgements

The authors would like to thank D. Kartashov and M. N. Slipchenko for valuable discussions. This work was supported by the U.S. Department of Energy (Grant No. DE SC0023209).

## References

[1] F. Krausz, and M. Ivanov, "Attosecond physics," Rev. Mod. Phys. **81**(1), 163 (2009).

[2] A. Couairon, and A. Mysyrowicz, "Femtosecond filamentation in transparent media," Phys.

Rep. **441**(2), 47 (2007).

[3] S.L. Chin, *Femtosecond Laser Filamentation* (Springer Science, 2010).

[4] P. Polynkin, and Y. Cheng, *Air Lasing* (Springer Series in Optical Sciences, 2018).

[5] N.L. Aleksandrov, S.B. Bodrov, M. V. Tsarev, A.A. Murzanev, Y.A. Sergeev, Y.A. Malkov, and A.N. Stepanov, "Decay of femtosecond laser-induced plasma filaments in air, nitrogen, and argon for atmospheric and subatmospheric pressures," Phys. Rev. E **94**(1), 013204 (2016).

[6] S. Bodrov, V. Bukin, M. Tsarev, A. Murzanev, S. Garnov, N. Aleksandrov, and A. Stepanov, "Plasma filament investigation by transverse optical interferometry and terahertz scattering," Opt. Express **19**(7), 6829 (2011).

[7] A.A. Ovsyannikov, and M.F. Zhukov, *Plasma Diagnostics* (Cambridge International Science Publishing, Cambridge, 2000).

[8] A. Talebpour, S. Larochelle, and S.L. Chin, "Dissociative ionization of NO in an intense laser field: A route towards enhanced ionization," J. Phys. B At. Mol. Opt. Phys. **30**(8), 1927 (1997).

[9] A. Talebpour, J. Yang, and S.L. Chin, "Semi-empirical model for the rate of tunnel ionization of N2 and O2 molecule in an intense Ti:sapphire laser pulse," Opt. Commun. **163**, 29 (1999).

[10] P. Polynkin, "Mobilities of O 2 + and O 2 - ions in femtosecond laser filaments in air," Appl. Phys. Lett. **101**(16), 164102 (2012).

[11] A.A. Ionin, S.I. Kudryashov, A.O. Levchenko, L. V. Seleznev, A. V. Shutov, D. V. Sinitsyn, I. V. Smetanin, N.N. Ustinovsky, and V.D. Zvorykin, "Triggering and guiding electric discharge by a train of ultraviolet picosecond pulses combined with a long ultraviolet pulse," Appl. Phys. Lett. **100**(10), 104105 (2012).

[12] R.P. Fischer, A.C. Ting, D.F. Gordon, R.F. Fernsler, G.P. DiComo, and P. Sprangle, "Conductivity measurements of femtosecond laser-plasma filaments," IEEE Trans. Plasma Sci. **35**(5 II), 1430 (2007).

[13] Y.H. Chen, S. Varma, T.M. Antonsen, and H.M. Milchberg, "Direct measurement of the electron density of extended femtosecond laser pulse-induced filaments," Phys. Rev. Lett. **105**(21), 215005 (2010).

[14] J. Papeer, C. Mitchell, J. Penano, Y. Ehrlich, P. Sprangle, and A. Zigler, "Microwave diagnostics of femtosecond laser-generated plasma filaments," Appl. Phys. Lett. **99**(14), 141503 (2011).

[15] A. V. Mitrofanov, A.A. Voronin, D.A. Sidorov-Biryukov, A. Puɟlys, E.A. Stepanov, G.

Andriukaitis, T. Flöry, S. Ališauskas, A.B. Fedotov, A. Baltuška, and A.M. Zheltikov, "Mid-infrared laser filaments in the atmosphere," Sci. Rep. **5**, 8368 (2015).

[16] A. V Mitrofanov, D.A. Sidorov-Biryukov, A.A. Voronin, A. Pugžlys, G. Andriukaitis, E.A. Stepanov, S. Ališauskas, T. Flöri, A.B. Fedotov, V.Y. Panchenko, A. Baltuška, and A.M. Zheltikov, "Subterawatt femtosecond pulses in the mid-infrared range: new spatiotemporal dynamics of high-power electromagnetic fields," Physics-Uspekhi **58**(1), 89–94 (2015).

[17] V. Shumakova, Filamentation and Self-Compression of Multi-MJ Fs Mid-Infrared Pulses, TU Wien, 2018.

[18] A. Patel, C. Gollner, R. Jutas, V. Shumakova, M.N. Shneider, A. Pugzlys, A. Baltuska, and A. Shashurin, "Ionization rate and plasma dynamics at 3.9 micron femtosecond photoionization of air," Phys. Rev. E **106**(5), 055210 (2022).

[19] J.F. Daigle, A. Jaroń-Becker, S. Hosseini, T.J. Wang, Y. Kamali, G. Roy, A. Becker, and S.L. Chin, "Intensity clamping measurement of laser filaments in air at 400 and 800 nm," Phys. Rev. A - At. Mol. Opt. Phys. **82**(2), 023405 (2010).

[20] J. Kasparian, R. Sauerbrey, and S.L. Chin, "The critical laser intensity of self-guided light filaments in air," Appl. Phys. B Lasers Opt. **71**(6), 877–879 (2000).

[21] A.R. Patel, A. Ranjan, X. Wang, M.N. Slipchenko, M.N. Shneider, and A. Shashurin, "Thomson and collisional regimes of in-phase coherent microwave scattering off gaseous microplasmas," Sci. Rep. **11**(1), 23389 (2021).

[22] A.R. Patel, Constructive (Coherent) Elastic Microwave Scattering-Based Plasma Diagnostics and Applications to Photoionization, PhD Dissertation, Purdue University, 2022.

[23] V. V. Semak, and M.N. Shneider, "Effect of power losses on self-focusing of high-intensity laser beam in gases," J. Phys. D. Appl. Phys. **46**(18), 185502 (2013).

[24] L.A. Finney, P.J. Skrodzki, M. Burger, X. Xiao, J. Nees, and I. Jovanovic, "Optical emission from ultrafast laser filament-produced air plasmas in the multiple filament regime," Opt. Express **26**(22), 29110–29122 (2018).

[25] C.J. Peters, M.N. Shneider, and R.B. Miles, "Kinetics model of femtosecond laser ionization in nitrogen and comparison to experiment," J. Appl. Phys. **125**(24), 243301 (2019).

[26] A. Shashurin, M.N. Shneider, A. Dogariu, R.B. Miles, and M. Keidar, "Temporally-resolved measurement of electron density in small atmospheric plasmas," Appl. Phys. Lett. **96**(17), 171502 (2010).

[27] A. Sharma, M.N. Slipchenko, M.N. Shneider, X. Wang, K.A. Rahman, and A. Shashurin, "Counting the electrons in a multiphoton ionization by elastic scattering of microwaves," Sci. Rep. **8**(1), 2874 (2018).

[28] A. Shashurin, A.R. Patel, X. Wang, A. Sharma, and A. Ranjan, "Coherent microwave scattering for diagnostics of small plasma objects: A review," Phys. Plasmas **30**(6), 063508 (2023).

[29] T.C. Murphy, *Total and Differential Electron Collision Cross Sections for 02 and N2 Reproduction* (Los Alamos, NM, 1988).

[30] Y.P. Raizer, *Gas Discharge Physics* (Springer, Berlin, 1991).

[31] I.A. Kossyi, A.Y. Kostinsky, A.A. Matveyev, and V.P. Silakov, "Kinetic scheme of the non-equilibrium discharge in nitrogen-oxygen mixtures," Plasma Sources Sci. Technol. **1**(3), 207–220 (1992).

[32] A. Dogariu, M.N. Shneider, and R.B. Miles, "Versatile radar measurement of the electron loss rate in air," Appl. Phys. Lett. **103**(22), 224102 (2013).

[33] V. Shumakova, E. Schubert, T. Balčiūnas, M. Matthews, S. Ališauskas, D. Mongin, A. Pugžlys, J. Kasparian, A. Baltuška, and J.-P. Wolf, "Laser induced aerosol formation mediated by resonant excitation of volatile organic compounds," Optica **8**(10), 1256 (2021).

[34] V. Shumakova, A. Pugžlys, S. Ališauskas, A. Baltuška, A. Voronin, A. V. Mitrofanov, D.A. Sidorov-Biryukov, A.M. Zheltikov, and D. Kartashov, in *2017 Conf. Lasers Electro-Optics, CLEO 2017 - Proc.* (2017), p. FM1F.5.